\documentclass[showpacs,twocolumn]{revtex4}
\usepackage{amsfonts}

\begin{document}

\title{Exponential spreading and singular behavior of quantum dynamics \\
near hyperbolic points }

\author{A. Iomin}
\affiliation{Department of Physics, Technion, Haifa, 32000,
Israel}

\begin{abstract}
Quantum dynamics of a particle in the vicinity of a hyperbolic
point is considered. Expectation values of dynamical variables are
calculated, and the singular behavior is analyzed. Exponentially
fast extension of quantum dynamics is obtained, and conditions for
this realization are analyzed.
\end{abstract}


\maketitle


Hyperbolic (saddle) points are a source of instability in
dynamical systems \cite{mackay}. Therefore, quantum dynamics of a
particle in a saddle point potential is an important problem in
quantum chaos \cite{bz,bery,gutzwiller}. The hyperbolic point at
the origin $(x,p)=(0,0)$ can be described locally by the
Hamiltonian $H_{loc}=xp$. The Lyapunov exponents detecting stable
$\Lambda_{-}$ and unstable $\Lambda_{+}$ manifolds are
$\Lambda_{\pm} =\pm 1$. This leads to the exponential spreading in
quantum dynamics and exponential growth of observable quantities
\cite{barton,quarneri}. The Hamiltonian $H_{loc}$ has been studied
in connection with the Riemann hypothesis \cite{berry,armitage},
scattering of the inverted harmonic oscillator \cite{bhaduri}, and
eigenstates near a hyperbolic point \cite{nonnemacher}.

We consider a quantum system which is a second order polynomial of
$H_{loc}$ with the Hamiltonian
\begin{eqnarray}\label{bv1}
\hat{H}_{s} &=&
2\omega\hat{x}\hat{p}-i\omega\hbar+\mu(2i\hat{x}\hat{p}+\hbar)^2
\nonumber \\
&\equiv& 2\omega[\hat{H}_{loc}-\frac{i\hbar}{2}]-
4\mu[\hat{H}_{loc}-\frac{i\hbar}{2}]^2 \, ,
\end{eqnarray}
where $\omega$ and $\mu$ are the linearity and nonlinearity
parameters correspondingly, while the coordinate and momentum
operators obey the standard commutation rule
$[\hat{x},\hat{p}]=i\hbar$ with Planck's constant $\hbar$. This
system was considered in \cite{armitage} in connection with the
Riemann hypothesis as well.  Our primary interest in this
Hamiltonian is related to a problem of quantum dynamics considered
in the Heisenberg picture \cite{ber_vish} where the expectation
values of the operators $\hat{p}(t)$ and $\hat{x}(t)$ were
calculated in the coherent states \cite{glauber}, specially
prepared at the initial moment $t=0$. As shown in \cite{ber_vish},
the dynamics of these expectation values becomes singular at
specific singularity times $t_l$. For example, for the observable
value of $\hat{x}^2$ these singularities occur at times $t_l=
\frac{\pi}{32\hbar\mu}+l\frac{\pi}{16\hbar\mu}, ~ l=0,\pm 1,\pm
2,\dots $. A remarkable property of these explosions is a pure
quantum nature: in the classical counterpart this corresponds to
the separatrix  motion without any singularities. As follows from
the analysis of Ref. \cite{ber_vish}, this explosion behavior
results from the interplay between the nonlinear term and the
specific choice of zero boundary conditions on infinities. When
$\mu=0$, the singularities are shifted to infinity and expectation
values of operators are finite, and this result is independent of
the boundary conditions.

We develop a different consideration of the problem to understand
the nature of this singularities that, as will be shown, are
related to the choice of the initial conditions. First we consider
the quantum dynamics of a particle of a unit mass in the saddle
potential described by the Hamiltonian (\ref{bv1}). For
calculation of the expectation values, following \cite{ber_vish},
the initial wave function is chosen in the form of the coherent
state $\Psi_0(x)=\langle x|\alpha\rangle$. In the
$x$-representation this is the Gaussian packet
\cite{glauber,ber_vish}:
\begin{equation}\label{bv10}
\Psi_0(x)=\langle x|\alpha\rangle=(\hbar\pi)^{-1/4}
e^{-|\alpha|^2/2\hbar
-\left(x^2-2\sqrt{2}x\alpha+\alpha^2\right)/2\hbar} \, ,
\end{equation}
where $\alpha\in \mathbb{C}$. It is worth mentioning that in this
notation the dimension of $x$ is $\sqrt{\hbar}$. For simplicity we
calculate the expectation value of the operator $\hat{x}^2(t)$.
Thus we have
\begin{equation}\label{bv11}
\langle\hat{x}^2(t)\rangle=\int_{-\infty}^{\infty}
\Psi_0^*(x)\hat{U}^{\dag}(t)x^2\hat{U}(t)\Psi_0(x)dx \, ,
\end{equation}
where $\hat{U}(t)$ is the evolution operator.

The axis of integration we split into three intervals
$(-\infty,-x_0],~[-x_0, x_0],~[x_0,\infty)$, and the expectation
value is expressed by the following three integrals:
\begin{eqnarray}\label{bv12}
&\langle\hat{x}^2(t)\rangle=I_{s}(t)+I_{f}^{-}(t)+I_{f}^{+}(t)
=\int_{-x_0}^{x_0}\Psi_{s}^*(t)x^2\Psi_{s}(t)dx
\nonumber \\
&+\int_{-\infty}^{-x_0}\Psi_{f}^{*}(t)x^2\Psi_{f}(t)dx+
\int_{x_0}^{\infty}\Psi_{f}^{*}(t)x^2\Psi_{f}(t)dx \, .
\end{eqnarray}
The dynamics in the finite interval $[-x_0, x_0]$ is considered in
the framework of the truncated interaction. Near the hyperbolic
point this dynamics is considered locally, such that $H=H_s$ for
$x<x_0$ and the particle is, for example, free with $H=H_f$
outside the interaction region $|x|>x_0$, where $ x_0>0$
determines arbitrary the interaction range. Note that we do not
consider a scattering task and just truncate the integration. Here
$\hat{H}_{s}$ is determined by Eq. (\ref{bv1}), while
$\hat{H}_{f}=\hat{p}^2/2$ determines free motion. Therefore, the
dynamics of an initial wave function $\Psi_0$ is determined in
these two different regions
\begin{equation}  \label{bv3}
\Psi_{f}(t)=\hat{U}_{f}(t)\Psi_0 ~~~\mbox{and}~~~
\Psi_{s}(t)=\hat{U}_{s}(t)\Psi_0 \, ,
\end{equation}
where the evolution operators $\hat{U}_{f}(t)=
\exp\left[-\frac{i}{\hbar}\hat{H}_{f}t\right]$ and
$\hat{U}_{s}(t)=\exp\left[-\frac{i}{\hbar}\hat{H}_{s}t\right]$
describe two independent processes, and corresponding shift of the
wave functions is supposed.

Integrals are calculated by substituting Eqs. (\ref{bv3}) in Eq.
(\ref{bv12}) and taking into account the explicit form of the
Hamiltonians. In the free motion window the evolution of the
square coordinate operator is
\begin{equation}\label{bv13}
\hat{x}^2(t)=\left[\hat{U}_{f}^{\dag}(t)\hat{x}\hat{U}_{f}(t)\right]^2=
\left[\hat{x}+t\hat{p}\right]^2 \, .
\end{equation}
Therefore, the free motion integrals $I_{f}^{\pm}(t)$ do not have
any particular features. Their values can be expressed in the form
of the error function
$\Phi(z)=(2\pi)^{-1/2}\int_{-\infty}^ze^{\eta^2/2}d\eta$
\cite{yel}. Then we obtain  that $I_{f}^{\pm}(t)\sim t^2$, as
expected.

The saddle point integral possesses a more interesting behavior.
Using calculations performed in \cite{ber_vish}[Eq. (4.16)], we
arrived at the following expression:
\begin{eqnarray}\label{bv14}
&I_{s}(t)=(\hbar\pi)^{-1/2}e^{4\omega t+24i\hbar\mu
t}e^{-\frac{(\alpha+\alpha^*)^2}{2\hbar}}
\int_{-x_0}^{x_0}dx\,x^2 \nonumber \\
&\times\exp\left\{ -\frac{x^2}{2\hbar}\left[\left(1+e^{32i\hbar\mu
t}\right)- 2\sqrt{2}\left(\alpha^*+\alpha e^{16i\hbar\mu
t}\right)x\right]\right\} \, .
\end{eqnarray}
First, we admit the exponential quantum growth, obtained in Ref.
\cite{ber_vish}. This quantum behavior has classical nature of the
near separatrix motion, observed also for a kicked system
\cite{quarneri}. At the  singularity times
$t_l=\frac{\pi}{32\hbar\mu}+\frac{l\pi}{16\hbar\mu}$  expression
(\ref{bv14}) is simplified, and the integral is calculated
exactly. It reads
\begin{eqnarray}\label{bv15}
&I_{s}(t_l)=\frac{(-i)^l}{(\hbar\pi)^{1/2}}e^{-\frac{(\alpha+\alpha^*)^2}
{2\hbar}}e^{\omega\pi/8\hbar\mu+l\omega\pi/4\hbar\mu}e^{3i\pi/4}
\nonumber \\
&\times \left[
\frac{\sqrt{2}\hbar^3}{\xi^3}\sinh\left(\frac{\sqrt{2}\xi
x_0}{\hbar} \right) - \frac{2\hbar^2
x_0}{\xi^2}\cosh\left(\frac{\sqrt{2}\xi x_0}{\hbar}
\right)\right. \nonumber \\
&-\left.\frac{\sqrt{2}\hbar
x_0^2}{\xi}\sinh\left(\frac{\sqrt{2}\xi x_0}{\hbar}
\right)\right]\, ,
\end{eqnarray}
where $\xi=\alpha^*+i(-1)^l\alpha$. This expression is finite for
finite $x_0$. When $x_0$ approaches infinity the integral diverges
and $t_l$ are the singularity points. Note that $x_0$ is an
arbitrary defined scale \cite{app1}.

To generalize the consideration of the explosion singularities,
first we consider eigenvalue problem for the Hamiltonian $H_s$.
Since the operator $\hat{x}\hat{p} -i\hbar/2$ commutes with
$\hat{H}_{s}$, this problem is reduced to the dimensionless
equation for the eigenfunctions $\chi_{\epsilon}(x)$:
\begin{equation}\label{bv16}
\frac{1}{i}\left(x\frac{d}{dx}+\frac{1}{2}\right)\chi_{\epsilon}(x)
=\epsilon\chi_{\epsilon}(x)
\end{equation}
with the following solution
\begin{equation}\label{bv17}
\chi_{\epsilon}(x)=\frac{1}{\sqrt{N_s|x|}}\exp[i\epsilon\ln|x|] \,
,
\end{equation}
which satisfies the boundary conditions
$\chi_{\epsilon}(x=\pm\infty)=0$ and $ N_{s}=4\pi\hbar^{1/2}$
\cite{app2}. For the continuous spectrum the normalization
condition is
\begin{equation}\label{bv18}
\int_{-\infty}^{\infty}\chi_{\epsilon'}^*(x)\chi_{\epsilon}(x)dx
=\delta(\epsilon-\epsilon')
\end{equation}
(see {\em e.g.} \cite{LL}).

Now, expanding  $\Psi_0(x)$ over the complete set of
$\chi_{\epsilon}(x)$,  we obtain
\begin{equation}\label{bv13a}
\Psi_0(x)=\int d\epsilon q(\epsilon)\chi_{\epsilon}(x) \, .
\end{equation}
Note, that the explicit form of the expansion coefficients
$q(\epsilon)$ is not important, since integration over energy
$\epsilon$ will be performed with exactly the same form of
$\chi_{\epsilon}(x)$. Substituting Eq. (\ref{bv13a}) in the
integral $I_{s}$ in Eq. (\ref{bv12}) with $x_0=\infty$, we obtain
\begin{eqnarray}\label{bv13b}
I_{s}^{\infty}(t)&=&2\int_{0}^{\infty}x^2\int
'q^*(\epsilon')q(\epsilon) e{-i(E-E')t}d\epsilon d\epsilon' \nonumber \\
&\times& \chi_{\epsilon'}^*(x) \chi_{\epsilon}(x)dx \, ,
\end{eqnarray}
where $E=2\omega\epsilon-4\hbar\mu\epsilon^2$ is the energy of
$\hat{H}_s$, and we use that
$\chi_{\epsilon}(-x)=\chi_{\epsilon}(x)$. The complex Gaussian
exponents are presented in the form of the Fourier integrals:
\begin{equation}\label{bv13c}
e^{\pm i4\hbar\mu t\epsilon^2}=\int_{-\infty}^{\infty}\frac{e^{\mp
i\tau\epsilon}d\tau}{\sqrt{\pm 16\pi i\hbar\mu t}}\exp\{\mp
i\tau^2/16\hbar\mu t\}  \, .
\end{equation}
Substituting these expressions in Eq. (\ref{bv13b}) and taking
into account the explicit form of $\chi_{\epsilon}(x)$, we obtain
$\ln |x| -2\omega t-\tau=\ln(|x|e^{-2\omega t}e^{-\tau})$. Then
one integrates over $\epsilon$ and $ \epsilon'$ to obtain the
following expression
\begin{eqnarray}\label{bv13d}
I_{s}(t)&=&2\int_{0}^{\infty}x^2 \int \frac{d\tau
d\tau'}{16\pi\hbar\mu t} e^{-i\frac{(\tau^2-{\tau'}^2)}{16\hbar\mu
t}}
e^{-2\omega t}e^{\frac{(\tau+\tau')}{2}} \nonumber \\
&\times&\Psi_0^*\left(xe^{-2\omega t}e^{-\tau'}\right)
\Psi_0\left(xe^{-2\omega t}e^{-\tau}\right)dx \, .
\end{eqnarray}
The next step is integration over $\tau$ and $\tau'$. To this end
we perform the following variables change $\tau=u+v$ and
$\tau'=u-v$ with the Jacobian of the transformation equaling 2.
Denoting $y=xe^{-2\omega t}e^{-(\tau+\tau')/2}$, integration over
$u$ is exact and gives the $\delta$ function $\delta(v-8i\hbar\mu
t)$. Therefore integration over $v$ is also exact. Finally, the
expectation value reads
\begin{equation}\label{bv13e}
\langle\hat{x}^2(t)\rangle = 2e^{4\omega
t}\int_{0}^{\infty}y^2\Psi_{0}^*\left(ye^{8i\hbar\mu t}\right)
\Psi_{0}\left(ye^{-8i\hbar\mu t}\right)dy \, .
\end{equation}
Here we also use the symmetrical property of the wave function.
Substituting Eq. (\ref{bv10}) in Eq. (\ref{bv13e}), one obtains
that at times $t=t_l$ this expression diverges. These are the same
singularities obtained above in Eq. (\ref{bv14}). Moreover, any
``good'' Gaussian and exponential functions lead to these kind of
singular behavior for the expectation values of physical
operators.

Obviously, these singularities result from a specific preparation
of the initial wave packets $\Psi_0(x)$. Let us prepare the
initial conditions ``properly'' to obtain the finite moments of
the physical variables. Owing to Eq. (\ref{bv13a}), we present the
initial condition as the spectral decomposition with the Gaussian
weight
$q(\epsilon)=\left[\frac{2a}{\pi}\right]^{\frac{1}{4}}\exp(-a\epsilon^2)$,
where real $a>0$. This yields the initial wave packet in the form
of a log-normal distribution
\begin{equation}\label{sd1}
\Psi_0(x)=\frac{1}{\sqrt{N}}
\exp\left(-\frac{1}{4a^2}\ln^2|x|-\frac{1}{2}\ln|x|\right)\,
\end{equation}
with $N=4a\sqrt{\pi}$. Using Eqs. (\ref{sd1}) and (\ref{bv13e}),
one obtains for the $n$th moment of $\hat{x}$
\begin{eqnarray}\label{sd2}
&{}&\langle\hat{x}^n(t)\rangle= \frac{1}{2a\sqrt{\pi}}e^{4\omega
t}e^{\frac{16}{a^2}\hbar^2\mu^2t^2} \nonumber \\
&\times& \int_{0}^{\infty}\exp\left(-\frac{1}{2a^2}\ln^2x+n\ln
x\right)d(\ln x)
\nonumber \\
&=&\frac{\sqrt{2}}{2}\exp\left(4\omega
t+a^4n^2+\frac{16}{a^2}\hbar^2\mu^2t^2\right)\, .
\end{eqnarray}
This behavior of the expectation values is finite and spreads
exponentially for the arbitrary long time scale. This exponential
increasing with time consists of two values. The first one is the
above mentioned classical term $e^{4\omega t}$, which is due to
the classical motion near the hyperbolic point. The second, pure
quantum, term $\frac{16}{a^2}\hbar^2\mu^2t^2$ is dominant and
relates to the action of the evolution operator, which is a
delation operator $e^{b\hat{xp}}$, where
$e^{b\hat{xp}}f(x)=f\left(e^{-i\sqrt{\hbar}b}\right)$ (see
\textit{e.g.}, \cite{teschl}). Therefore, this term is due to the
quantum dynamics near the hyperbolic point.

Both singular behavior and this, completely new, exponential
quantum growrth are due to the nonlinear quantum dispersion term
$\kappa=\hbar\mu T$, where $T$ is a characteristic time scale. For
example, for the explosion behavior of
$\langle\hat{x}^2(t)\rangle$ in  Eqs. (\ref{bv14}) and
(\ref{bv15}) it is $T=\frac{\pi}{32\hbar\mu}$. For the quantum
expansion we can define this time scale parameter as
$\hbar\mu/\omega$ to define the dimensionless growth of the
expectation values
$\exp\left[\left(\frac{\kappa}{a}\right)^2\omega t\right]$.

This research was supported by the Israel Science Foundation.


\begin{thebibliography}{99}
\bibitem{app1} The interaction range $x_0$ is always a
finite value due to the local consideration near the hyperbolic
point. For example, it can be evaluated as the difference between
normalization constants of the wave functions for the free motion
$N_{f}$ and for the saddle point motion $N_{s}$ (see, Problem
133.1 in \cite{LL}). For the case of free motion the normalization
constant $N_{f}$ is determined from the integral \cite{LL,fok} \\
$\int_{-\infty}^{\infty}e^{-ipx/\hbar}e^{ip'x/\hbar}dx=
2\pi\sqrt{\hbar}\delta(p-p')\equiv N_f\delta(p-p')\, . $ \\
Therefore, $N_{f}=2\pi\hbar^{1/2}$. We also used here that $x$ has
the dimension $\hbar^{1/2}$.  In order to find the $N_{s}$, one
uses the eigenfunction of the Hamiltonian $\hat{H}_{s}$ with an
extended action on the whole $x$ axis from minus to plus infinity.
Eventually we obtain  that the interaction range is
$x_0=N_s-N_f=2\pi\hbar^{1/2}$.
\bibitem{app2} A mathematically rigorous calculation of the
normalization constant for the wave function $\chi_{\epsilon}(x)$
can be pressented by follow the presentation in the monograph by
V.A. Fock \cite{fok}. Since the operator $\hat{x}\hat{p}$ has
continuous spectrum  $\epsilon$, the eigenfunctions
$\chi(\epsilon,x)\equiv\chi_{\epsilon}(x)$ are not square
integrable. Therefore, the normalization condition exists not for
the eigenfunction but for the ``eigendifferential'' \cite{fok}
$\Delta\chi(\epsilon,x)$, which reads

$ \Delta\chi(\epsilon,x)=\int_{\epsilon}^{\epsilon+\Delta\epsilon}
\chi(\epsilon',x)d\epsilon' \, .$

\noindent Substituting here Eq. (\ref{bv17}), one obtains

$ \Delta\chi(\epsilon,x)=\frac{2}{\sqrt{N|x|}\ln|x|}
\exp[i(\epsilon+\Delta\epsilon)\ln|x|]\sin\frac{\Delta\epsilon\ln|x|}{2}
\, .$

\noindent This solution is already square integrable and has the
following normalization form

$\lim_{\Delta\epsilon\to 0}\frac{1}{\Delta\epsilon}
\int_{-\infty}^{\infty}dx\left|\Delta\chi(\epsilon,x)\right|^2=\sqrt{\hbar}
\, .$

\noindent Carrying out the following variable change
$z=(\Delta\epsilon/2)\ln |x|$ and taking into account that $
\int_{-\infty}^{\infty}\frac{\sin^2z}{z^2}dz=\pi \, ,$ we obtain
$N=4\pi$ that coincides exactly with the dimensionless
normalization constant $N_s\hbar^{1/2}$ in Eq. (\ref{bv17}).
\end{thebibliography}
\end{document}